\documentclass[prd,nofootinbib,english]{revtex4}
\usepackage{amsmath}
\usepackage{amsfonts,color}
\usepackage{amssymb,float,multirow}
\usepackage[utf8]{inputenc}
\usepackage{color}
\usepackage[unicode=true,pdfusetitle,
bookmarks=true,bookmarksnumbered=true,bookmarksopen=true,bookmarksopenlevel=3,
breaklinks=false,pdfborder={0 0 1},backref=false,colorlinks=true]
{hyperref}
\usepackage{enumitem}

\usepackage{tikz}
\usetikzlibrary{shapes.geometric}
\usetikzlibrary{arrows.meta,arrows}

\begin{document}

\title{Exact spherically symmetric solutions in modified Gauss-Bonnet gravity from Noether symmetry approach}

\author{Sebastian Bahamonde}\email{sbahamonde@ut.ee, sebastian.beltran.14@ucl.ac.uk}
\affiliation{Laboratory of Theoretical Physics, Institute of Physics, University of Tartu, W. Ostwaldi 1, 50411 Tartu, Estonia}
\affiliation{Department of Mathematics, University College London, Gower Street, London, WC1E 6BT, United Kingdom}
	\author{Konstantinos Dialektopoulos}\email{kdialekt@gmail.com}\affiliation{Center for Gravitation and Cosmology, College of Physical Science and Technology, Yangzhou University, Yangzhou 225009, China}
\author{Ugur Camci}\email{ugurcamci@gmail.com}
\affiliation{Siteler Mahallesi, 1307 Sokak, Ahmet Kartal Konutlari, A-1
Blok, No:7/2, 07070, Konyaalti, Antalya, Turkey}

\begin{abstract}
It is broadly known that Lie point symmetries and their subcase, Noether symmetries, can be used as a geometric criterion to select alternative theories of gravity. Here, we use Noether symmetries as a selection criterion to distinguish those models of $f(R,G)$ theory, with $R$ and $G$ being the Ricci and the Gauss-Bonnet scalars respectively, that are invariant under point transformations in a spherically symmetric background. In total, we find ten different forms of $f$ that present symmetries and calculate their invariant quantities, i.e Noether vector fields. Furthermore, we use these Noether symmetries to find exact spherically symmetric solutions in some of the models of $f(R,G)$ theory. 
\end{abstract}


%\date{\today}

\maketitle
%\tableofcontents

%%%%%%%%%%%%%%%%%%%%%%%%%%%
%%%  Sec. I
%%%%%%%%%%%%%%%%%%%%%%%%%%%	
\section{Introduction}\label{sec:0}
General Relativity (GR) is known to be the most successful theory for the gravitational interactions so far. However, it possesses some shortcomings that have led many scientists to pursue an alternative theory for the description of gravity. These are mostly related to the \textit{dark universe}, i.e. the nature of dark energy and dark matter, as well as the inability to find a TeV scale supersymmetry, the nature of singularities, the value of the cosmological constant and other less important astrophysical problems~\cite{Wang2007,Wang2008}. 

Based on these, numerous modifications of gravity have been proposed~\cite{Clifton:2011jh,Capozziello:2011et,Nojiri:2017ncd}. The addition of a scalar field, originally proposed by Brans and Dicke~\cite{Brans:1961sx} in the early sixties, adds a new scalar degree of freedom to the theory and gravity is no longer uniquely described from the metric. The most general scalar tensor theory with a single scalar field, leading to second order field equations is the Horndeski theory~\cite{Horndeski:1974wa}. After the observation of gravitational waves though, a significant part of it is severely constrained from the speed of the graviton. Another significant modification (or better \textit{extension}) of gravity, is the so called $f(R)$ gravity~\cite{Capozziello:2011et,Sotiriou:2008rp}, which generalizes the Einstein-Hilbert action to and arbitrary function of the Ricci scalar $R$. Many models of $f(R)$ gravity have been studied in different contexts in the literature. As appealing as they may be, most of them have partial success in specific scales; some give a very good description of the early universe, but fail to describe late-time acceleration \cite{Starobinsky:1980te}, others succeed in describing today's cosmology but lack in describing earlier epochs \cite{Hu:2007nk}, others give a convincing explanation of the rotation curves of galaxies, without invoking exotic (i.e. dark) forms of matter \cite{Capozziello:2012ie,Boehmer:2007kx}, and so on. Last but not least, many modifications have been also studied in the so-called teleparallel geometry. There, the curvature of the manifold vanishes and gravity is purely a torsional effect; the gravitational field is described by the tetrad (vierbein) and the spin connection of the tangent space. The interested reader should read~\cite{Aldrovandi:2013wha,Krssak:2018ywd,Bahamonde:2019shr,Bahamonde:2015zma,Bahamonde:2017wwk,Bahamonde:2018miw}.

All the above theories mentioned are motivated from  cosmology. There are many modifications of gravity that find their origins in high energies, such as Kaluza-Klein \cite{Coquereaux:1990qs}, Dvali-Gabadadze-Porrati  (DGP)\cite{Dvali:2000hr}, Einstein-Dilaton-Gauss-Bonnet \cite{Lovelock:1971yv}, or Randall-Sundrum \cite{Randall:1999ee}.

A topological invariant in $3+1$ dimensions appears in the low energy limit of string theories, that is known as Gauss-Bonnet scalar \cite{Gasperini:1992em} and is given by
\begin{equation}
\label{GBdef}
G= R^2 - 4 R_{\mu\nu}R^{\mu\nu} + R_{\alpha\beta\mu\nu}R^{\alpha\beta\mu\nu}\,,
\end{equation}
with $R$, $R_{\mu\nu}$ and $R_{\alpha\beta\mu\nu}$ being the Ricci scalar, the Ricci tensor and the Riemann tensor respectively. Being a topological invariant, a linear term of it in a four-dimensional action would contribute a total derivative which should vanish at infinity, yielding trivial equations of motion. It turns out though that when it appears coupled to either a scalar field or to the metric (and its derivatives) one gets significant phenomenology both in high energies and in cosmology~\cite{Carroll2005}.

It is well known that Noether symmetries (i.e. invariance under point transformations) can be used as a geometric criterion to discriminate between alternative theories of gravity~\cite{Dialektopoulos:2018qoe}. The existence of symmetries means that the dynamical system has some conserved quantities that, in many cases, are related to physical/observed quantities. Hence, this renders the possibility to use it as a theoretical constraint in order to classify different models, based on the geometric symmetries of the spacetime~\cite{Dialektopoulos:2018qoe}. In addition, these symmetries could also help us in reducing the dynamics of a system and thus find  exact solutions more easily. This method has been extensively used in the literature, both in cosmological and astrophysical scales~\cite{Bahamonde:2019jkf,Bahamonde:2018ibz,Bahamonde:2017sdo,Bahamonde:2016grb,Bahamonde:2016jqq,Capozziello:2007wc,Basilakos:2011rx,Basilakos:2013rua,Dialektopoulos:2019mtr,Capozziello:2018gms,Capozziello:2016eaz}.
Some cosmological paradigms such as quintessence and phantom ones can be recovered in the framework of $f(R,G)$ theories of gravity~\cite{Laurentis2015}. The successful realization of the dark energy and of the inflationary era is studied for some classes of $f(R,G)$ models~\cite{Odintsov2019}. Both the vacuum and the non-vacuum theories of $f(R,G)$ gravity admitting Noether symmetries are worked by 
\cite{Dialektopoulos:2018qoe,Capo2014,Ugur2018}, and they found some exact cosmological solutions. Obviously, apart from Noether symmetries, one could use contact symmetries, Cartan symmetries \cite{Karpathopoulos:2017arc}, etc to studying the invariance of a system of differential equations under specific types of transformations. 

In this paper, we will study the invariance of $f(R,G)$ theory of gravity, which is a generalization of $f(R)$ theory containing a Gauss-Bonnet scalar in the arbitrary function, under point transformations in a spherically symmetric background. We will present those models that possess Noether symmetries, calculate their invariant functions and also find exact spherical solutions in some of the cases.  

The paper is organized as follows: in Sec.~\ref{sec:II} we present the form of the theory $f(R,G)$, we write down its equations of motion and the form they take in a spherically symmetric spacetime. In Sec.~\ref{sec:III} we derive its point-like Lagrangian in the same spacetime and show that while the configuration space of it is five dimensional, i.e. $\mathcal{Q}=\{A,B,M,R,G\}$, its tangent space is only nine dimensional because there is no dependence on the derivative of the $B$ potential. This means that $B$ is a cyclic variable and one can solve for it and substitute its form back in the Lagrangian to obtain its canonical form. However, the Noether Symmetry Approach works even for non-canonical Lagrangians and for the sake of simplicity we keep it as it is. The next section, \ref{sec:IV}, contains a review of the Noether Symmetry Approach and the classification of $f(R,G)$ models that present such symmetries, together with the associated invariant quantities and solutions. We conclude our main results in Sec.~\ref{sec:V}.

%%%%%%%%%%%%%%%%%%%%%%%%%%%
%%%  Sec. II
%%%%%%%%%%%%%%%%%%%%%%%%%%%
\section{The $f(R,G)$ gravity in spherically symmetric space-time}
\label{sec:II}
Let us start from the most general action for modified Gauss-Bonnet gravity in 4-dimensions.
\begin{equation}
S=\int d^4x \sqrt{-g}\left[ \frac{1}{2\kappa^2} f(R,G) + L_{\rm m} \right]\,,
\label{action}
\end{equation}
where $\kappa^2=8\pi G$, $L_{\rm m}$ is any matter Lagrangian and $f(R,G)$ is a function which depends on the scalar curvature $R$ and the Gauss-Bonnet invariant $G$ being defined as in Eq.~\eqref{GBdef}. It is noted that any linear combination of the Gauss-Bonnet invariant does not
contribute to the effective Lagrangian in 4-dimension. Variations of the action \eqref{action} with respect to the metric tensor $g_{\mu \nu}$ yields
\begin{eqnarray}
& & f_{R} G_{\mu\nu}= \kappa^2 T_{\mu \nu} + \frac{1}{2}\left( f - R f_{R} - G f_G \right) g_{\mu\nu} + \nabla_\mu \nabla_\nu f_{R} - g_{\mu \nu} \square f_{R} + 2 R \nabla_{\mu} \nabla_{\nu} f_G  -2 g_{\mu \nu} R \square f_G  \nonumber \\& & \qquad \qquad - 4 R_{\mu}^{\alpha} \nabla_{\alpha} \nabla_{\nu} - 4 R_{\nu}^{\alpha} \nabla_{\alpha} \nabla_{\mu} + 4 R_{\mu \nu} \square f_G + 4 g_{\mu \nu} R^{\alpha \beta} \nabla_{\alpha} \nabla_{\beta} f_G + 4 R_{\mu \alpha \beta \nu} \nabla^{\alpha} \nabla^{\beta} f_G\,,
\label{fieldeq1}
\end{eqnarray}
where $f_R=\partial f/\partial R,$ $f_G=\partial f/\partial G$ and $\square$ is the d’Alembert operator in curved spacetime, respectively. Hereafter, we have assumed the vacuum case where $L_{\rm m}=0$. Let us now consider that the space-time is spherically symmetric such as the metric is
\begin{eqnarray}
ds^2=-A(r)dt^2 + B(r)dr^2 + M(r) (d\theta^2 + \sin\theta^2 d\varphi^2)\,, \label{metric}
\end{eqnarray}
where $A(r), B(r)$ and $M(r)$ are positive functions of the radial coordinate $r$. The scalar curvature and Gauss-Bonnet invariant for this space-time are 
\begin{eqnarray}
& & R = -\frac{1}{B} \left[ \frac{ 2M''}{M} + \frac{A''}{A} - \frac{A' B'}{2 A B} + \frac{M' A'}{M A} - \frac{M' B'}{M B} - \frac{A'^2}{2 A^2} - \frac{M'^2}{2 M^2} -  \frac{2 B}{M}  \right]\,, \label{r-scalar} \\ & & G= - \frac{2}{B M} \left( 2 \frac{A''}{A} - \frac{A' B'}{A B} - \frac{A'^2}{A^2} \right) + \frac{1}{B^2} \left[ \frac{M'^2}{M^2} \left( \frac{A''}{A} - \frac{3 A' B'}{2 A B} - \frac{ A' M'}{ A M} - \frac{A'^2}{2 A^2} \right) + 2 \frac{ A' M' M''}{ A M^2}  \right]\,. \label{gaussbonnet}
\end{eqnarray}
For the metric \eqref{metric}, the field equations \eqref{fieldeq1} become
\begin{eqnarray}
f_R \left( \frac{2 A''}{A} - \frac{A' B'}{A B} - \frac{A'^2}{A^2} + 2 \frac{M' A'}{M A} \right)& = & - 2 B ( f - G f_G) + 4 \left[ f''_R + f'_R \left( \frac{M'}{M} - \frac{B'}{2 B} \right)  \right] \nonumber \\& &  + \frac{4}{B} \left[  f''_G \left( \frac{4 B}{M} - \frac{M'^2}{M^2} \right) - f'_G \left\{ B' \left( \frac{2}{M} - \frac{3 M'^2}{2 B M^2} \right) + \frac{2 M' M''}{M^2} - \frac{M'^3}{M^3}  \right\}  \right] \,, \label{feq1}
\end{eqnarray}
\begin{equation}
f_R \left( \frac{2 A''}{A} + \frac{4 M''}{M} - \frac{A' B'}{A B} - \frac{2 M' B'}{M B} - \frac{A'^2}{A^2} -  \frac{2 M'^2}{M^2}  \right) = - 2 B ( f - G f_G) + 2 f'_R \left( \frac{A'}{A} +  \frac{2 M'}{M} \right) + \frac{2 A' f'_G }{A} \left( \frac{4}{M} - \frac{3 M'^2}{ B M^2} \right)\,,  \label{feq2}
\end{equation}
\begin{eqnarray}
f_R \left( \frac{2 M''}{M} - \frac{B' M'}{B M} +  \frac{M' A'}{M A} - \frac{4 B}{M} \right) & =& - 2 B ( f - G f_G) + 4 \left[ f''_R + \frac{1}{2} f'_R \left( \frac{A'}{A} -\frac{B'}{B} + \frac{M'}{M} \right) \right]   \nonumber \\& & - \frac{4}{B} \left[ f''_G \frac{M' A'}{M A} + f'_G \left\{  \frac{A' M''}{A M} + \frac{M'}{2 M} \left(  \frac{2 A''}{A}- \frac{A' M'}{A M} - \frac{3 A' B'}{A B} - \frac{A'^2}{A^2}  \right)   \right\}  \right]\,. \qquad \label{feq3}
\end{eqnarray}
In above equations, primes denote differentiation with respect to $r$, therefore, the terms $f'_{R}=f_{RR}R' + f_{RG} G'$ and $f'_{G}=f_{RG} R' + f_{GG} G' $.

%%%%%%%%%%%%%%%%%%%%%%%%%%%
%%%  Sec. III
%%%%%%%%%%%%%%%%%%%%%%%%%%%
\section{The point-like $f(R,G)$ Lagrangian}
\label{sec:III}
This section is devoted to find the point-like Lagrangian for $f(R,G)$ gravity in spherical symmetry. For simplicity let us express the scalar curvature $R$ and the Gauss-Bonnet invariant $G$ as follows
\begin{eqnarray}
& & \bar{R} = R^{*} - \frac{A''}{AB} - \frac{2M''}{B M}\,, \\& & \bar{G} = G^{*} - \frac{4}{B M} \frac{A''}{A} + \frac{M'}{B^2 M} \left( \frac{M' A''}{M A} + \frac{2 A' M''}{A M} \right)\,,
\end{eqnarray}
where $R^*$ and $G^*$
\begin{eqnarray}
& & R^{*}=\frac{A' B'}{2 A B^2} -\frac{A' M'}{A B M}+\frac{A'^2}{2 A^2 B}+\frac{B' M'}{B^2 M}+\frac{M'^2}{2 B M^2}+\frac{2}{M}\,, \\& & G^{*} = \frac{2}{B M} \left( \frac{A' B'}{A B} + \frac{A'^2}{A^2} \right) - \frac{M'}{B^2 M} \left( \frac{3 A' B' M'}{2 A B M} + \frac{M' A'^2}{2 M A^2} + \frac{A' M'^2}{A M^2} \right)\,,
\end{eqnarray}
contain only first derivatives terms. One can rewrite the action into its canonical form in such a way that we can reduce the number of degrees of freedom. In our case, we have
\begin{equation}
S_{f(R,G)}=\int dr\mathcal{L}(A,A',B,B',M,M',R,R',G,G')\,.  \label{actioncan0}
\end{equation}
Then, the action \eqref{action} in a spherically symmetric space-time \eqref{metric} becomes
\begin{equation}
S_{f(R,G)}= \int dr\left\{ f(R,G)-\lambda_{1} \left( R - \bar{R} \right) - \lambda_{2} \left( G - \bar{G} \right) \right\} M \sqrt{AB}\,. \label{actioncan}
\end{equation}
Here, $\lambda_1$ and $\lambda_2$ are the Lagrangian multipliers that can be directly found by varying with respect to $R$ and $G$, giving $\lambda_1=f_{R}$ and $\lambda_2=f_{G}$ respectively. Then, the above canonical action can be rewritten as
\begin{eqnarray}
S_{f(R,G)}& = & \int dr\Big\{ f(R,G)-f_{R} \left[ R - \Big(R^{*}-\frac{A''}{A B}-\frac{2M''}{B M} \Big) \right] \nonumber \\& &   -f_{G} \left[ G - \left( G^{*} - \frac{4}{B M} \frac{A''}{A} + \frac{M'^2 A''}{A B^2 M^2} + \frac{2 A' M' M''}{A B^2 M^2} \right) \right] \Big\} M \sqrt{AB} \,,  \label{actioncan2}\\
& = &  \int dr\Big\{ M \sqrt{AB} \Big[f(R,G)-f_{R}(R - R^{*}) - f_G ( G -G^{*} ) \Big] + 2 M' \Big(\sqrt{\frac{A}{B}}f_{R}\Big)' + A'\Big(\frac{Mf_R}{\sqrt{AB}}\Big)' \nonumber\\
&&  + 4 A' \left( \frac{ f_G}{\sqrt{A B}} \right)' - A' M'^2 \left( \frac{ f_{G} }{ M \sqrt{A B^3}} \right)' \Big\} \, ,  \label{actioncan3}
\end{eqnarray}
where we have integrated by parts and ignored boundary terms. Then, the point-like Lagrangian becomes
\begin{eqnarray}
& & \mathcal{L}_f =  f_R \left( \frac{ M' A'}{\sqrt{A B}} +  \sqrt{\frac{A}{B}}\frac{M'^2}{2 M} \right) + f'_R \left( \frac{M A' }{\sqrt{A B}}  + 2\sqrt{\frac{A}{B}}  M'  \right) +  \frac{f'_G}{ \sqrt{A B}}  \left( 4 A' - \frac{A' M'^2}{ M B} \right) \nonumber \\& &   + \sqrt{A B} \left[ M ( f - G f_G) + (2- M R) f_R \right]  .  \label{lagr}
\end{eqnarray}
Note again that $f_R'=f_{RR}R'+f_G G'$. 

Since the equation of motion (\ref{feq2}) describing the evolution of the metric potential $B$ does not depend on its derivative, it can be written as a quadratic equation in terms of $B$
\begin{equation}
\left[ f + \left( \frac{2}{M} -R \right) f_R - G f_G \right] B^2 + F(A,M,R,G) B - 3 f'_G \frac{A' M'^2}{A M^2} = 0 \,, \label{veq-2-2-1}
\end{equation}
which can be explicitly solved as a function of other coordinates for the roots of this quadratic equation such that
\begin{equation}
B_{1,2} = - \frac{F(A,M,R,G) }{f + \left( \frac{2}{M} -R \right) f_R - G f_G} \pm \frac{\sqrt{F(A,M,R,G)^2 + 12 f'_G \frac{A' M'^2}{A M^2} \left[ f + \left( \frac{2}{M} -R \right) f_R - G f_G \right]}}{2 \left[ f + \left( \frac{2}{M} -R \right) f_R - G f_G \right]} \,,  \label{veq-2-2}
\end{equation}
where $f + \left( \frac{2}{M} -R \right) f_R - G f_G \neq 0$ and $F(A,M,R,G)$ is defined as
\begin{equation}
F (A,M,R,G) \equiv  f_R \frac{M'}{M} \left( \frac{A'}{ A} + \frac{M'}{2 M} \right) + f'_R \left( \frac{A'}{A} + \frac{2 M'}{M} \right) + 4 f'_G \frac{A'}{M A}\, . \label{veq-2-2-2}
\end{equation}
The discriminant $\Delta$ is the part inside the square root in (\ref{veq-2-2}), where $\Delta \equiv F^2 + 12 f'_G \frac{A' M'^2}{A M^2} \left[ f + \left( \frac{2}{M} -R \right) f_R - G f_G \right]$, which comes from the quadratic formula and we can use this to find the nature of the roots. The roots of a quadratic equation with real coefficients are real and distinct if the discriminant is positive, are real with at least two equal if the discriminant is zero, and include a conjugate pair of complex roots if the discriminant is negative.
The discriminant can be used in the following way: there are no real roots if $\Delta <0$, the roots are real and equal, i.e. one real root if $\Delta = 0$ and the roots are real and unequal, i.e. two distinct real roots if $\Delta >0$. In this study, we discard the case $\Delta <0$ where the roots are not real. When there are two distinct real roots, then the condition $\Delta >0$ means that $f'_G \frac{A'}{A} \left[ f + \left( \frac{2}{M} -R \right) f_R - G f_G \right] >0$.

The energy functional $E_{\mathcal{L}}$ or the Hamiltonian of the Lagrangian $\mathcal{L}$  is defined by
\begin{eqnarray}
E_{\mathcal{L}} = q'^i \frac{\partial \mathcal{L}}{\partial q'^i} -\mathcal{L}\,. \label{energy}
\end{eqnarray}
Now, we calculate the energy functional $E_{\mathcal{L}_f}$ for the Lagrangian density $\mathcal{L}_f$ which has the form
\begin{eqnarray}
& & E_{\mathcal{L}_f} = M \sqrt{\frac{A}{B}} \Big\{ f_R \left( \frac{M' A'}{M A} +  \frac{M'^2 }{2 M^2} \right) + f'_R \left( \frac{A'}{A} + 2 \frac{M'}{M} \right)  + f'_G \frac{A'}{A B} \left( \frac{4 B}{M} - \frac{3 M'^2}{M^2} \right) \nonumber \\& &  - \frac{B}{M} \left[ M ( f - G f_G ) + (2 - MR) f_R  \right] \Big\} \,. \label{ef1}
\end{eqnarray}
Note that the energy function  $E_{\mathcal{L}_f}$ vanishes due to the field equation (\ref{feq2}) which is obtained by varying the Lagrangian (\ref{lagr}) according to the metric variable $B$. Therefore, the solution of equation $E_{\mathcal{L}_f} = 0$ in terms of $B$ is given by (\ref{veq-2-2}).

As it is expected due to the absence of the generalized velocity $B$ in the point-like Lagrangian (\ref{lagr}), the Hessian determinant of the Lagrangian (\ref{lagr}), which is defined by $\| \partial^2 \mathcal{L}_f /\partial q'^i \partial q'^j  \|$, is zero.  It is known that the metric variable $B$ does not contribute to the dynamics due to the point-like Lagrangian approach, but the equation of motion for $B$ has to be considered as a further constraint equation.

\bigskip

%%%%%%%%%%%%%%%%%%%%%%%%%%%
%%%  Sec. IV
%%%%%%%%%%%%%%%%%%%%%%%%%%%
\section{Noether symmetry approach}
\label{sec:IV}

In this section, we seek for the condition in order that the Lagrangian density (\ref{lagr}) would admit any Noether symmetry which has a generator of the form  %\cite{cimall}
\begin{equation}
{\bf X} = \xi \frac{\partial}{\partial r} + \eta^i \frac{\partial}{\partial q^i}\,, \label{ngs-gen}
\end{equation}
where $q^i$ are the generalized coordinates in the $d$-dimensional configuration space ${\cal Q }\equiv \{ q^i, i=1, \ldots, d \} $ of the Lagrangian, whose tangent space is ${\cal TQ }\equiv \{q^i,q'^i\}$. The components $\xi$ and $\eta^i$ of the Noether symmetry generator ${\bf X}$ are functions of $r$ and $q^i$. The existence of a Noether symmetry implies the existence of a vector field ${\bf X}$ given in (\ref{ngs-gen}) if the Lagrangian $ \mathcal{L}(r, A, B, M, R, G, A', B', M', R', G' )$ satisfies
\begin{equation}
{\bf X}^{[1]} \mathcal{L} + \mathcal{L} ( D_r \xi) = D_r K\, , \label{ngs-eq}
\end{equation}
where ${\bf X}^{[1]}$ is the first prolongation of the generator (\ref{ngs-gen}) in such a form
\begin{equation}
{\bf X}^{[1]} = {\bf X}  + \eta'^i \frac{\partial}{\partial q'^i}\,,
\end{equation}
and $K(r, q^i)$ is a gauge function, $D_r$ is the total derivative operator with respect to $r$, $D_r =\partial / \partial r + q'^i \partial / \partial q^i$, and $\eta'^i$ is defined as $\eta'^i = D_r \eta^i - q'^i D_r \xi$. The significance of Noether symmetry comes from the following first integral that if ${\bf X}$ is the Noether symmetry generator corresponding to the Lagrangian $\mathcal{L}(r, q^i, q'^i)$, then the Hamiltonian or a conserved quantity associated with the generator ${\bf X}$ is
\begin{equation}
I =- \xi E_{\mathcal{L}} + \eta^i \frac{\partial \mathcal{L}}{\partial q'^i} - K \,. \label{con-law}
\end{equation}

Let us start with the Lagrangian (\ref{lagr}), where $q^i = \{ A, B, M, R, G \}, i=1,\ldots, 5 $. Then the Noether symmetry condition (\ref{ngs-eq}) for this Lagrangian yields an overdetermined system of {\it 60}  partial differential equations. We will now solve  these differential equations, which will fix the symmetry generator ${\bf X}$ and the form of function $f(R,G)$ as in the following possibilities. There are several cases in which one of the quantities $f_{RG}, f_{RR}$ or $f_{GG}$ vanishes or not, such a way that {\bf (i)} $f_{RG} = 0$; {\bf (ii)} $f_{RR} = 0$; {\bf (iii)} $f_{GG} = 0$; and {\bf (iv)} $f_{RG}, f_{RR}, f_{GG} \neq 0$ (general case).

There is one Noether's vector that solve all the 60 differential equations that holds for any arbitrary function $f(R,G)$, that is given by
\begin{eqnarray}
& & {\bf X}_0 = \alpha (r) \partial_r - 2 B \alpha' (r) \partial_B\,, \label{X0-1}
\end{eqnarray}
where $\alpha (r)$ is an arbitrary function. Then, all particular theories will have ${\bf X}_0$ and also more Noether's vectors. Let us now split the study in different branches depending on the form of $f$.

\subsection{$f(R,G)=f_1(R) + f_2(G)$ ($f_{RG} = 0$)}

\subsubsection{Subcase: Pure $f(R)$ gravity ($f_{2,GG}(G)=0$)}
\begin{center}
I) General Relativity case: $f(R)=R$
\end{center}
 Here, we find {\it four} Noether symmetries, ${\bf X}_0$ and,
\begin{eqnarray}
& & {\bf X}_1 = - A \partial_A + B \partial_B + M \partial_M\,, \quad {\bf X}_2 = \frac{1}{\sqrt{M}} \partial_A - \frac{B}{A \sqrt{M}} \partial_B\,,  \label{X12-1} \\  & & {\bf X}_3 = -\frac{A}{2 \sqrt{M}} \partial_A + \frac{B}{2 \sqrt{M}} \partial_B + \sqrt{M} \partial_M\,. \label{X3-1}
\end{eqnarray}
The corresponding first integrals of these Noether symmetries give rise to $ E_{\mathcal{L}} = 0$ for ${\bf X}_0$, which means
\begin{equation}
B = \frac{ M' A'}{2 A} + \frac{M'^2}{4 M}\,, \label{NC0-1}
\end{equation}
and
\begin{eqnarray}
& & I_1 = \frac{ f_0 M A'}{\sqrt{A B}}\,, \qquad I_2 = \frac{ f_0 M'}{\sqrt{ A B M}}\,,  \qquad   I_3 = \frac{ I_1}{ \sqrt{M} }  + \frac{I_2}{2} A\,, \label{NC123-1}
\end{eqnarray}
for ${\bf X}_1, {\bf X}_2$ and ${\bf X}_3$. Then, the above equations \eqref{NC0-1} and \eqref{NC123-1} have the solutions
\begin{equation}
A(r) = \frac{2}{I_2} \left( I_3 - \frac{I_1}{\sqrt{M}}  \right)\,, \qquad B(r) = \frac{I_3}{2 I_2} \frac{ M'^2}{ M A}\,,  \label{sol-1}
\end{equation}
with $I_2 \neq 0$ and $I_3 = 2 f_0^2 / I_2$. If $M(r) = r^2$, the solutions \eqref{sol-1} become
\begin{equation}
A(r) =   A_0 - \frac{A_1}{r}\,, \qquad B(r) = \frac{A_0}{ A_0  - \frac{A_1}{r} }\,,  \label{sol-2}
\end{equation}
where $A_0 = 2 I_3 / I_2$ and $A_1 = 2 I_1 / I_2$. This is the well-known Schwarzschild solution for $A_0 = 1$ and $A_1 = 2 m$, where $m$ is the mass parameter. Then, we have easily found the Schwarzschild solution in GR using the first integrals.

\begin{center}
II) Power-law forms
\end{center}
It is also possible to find solutions to the system~\eqref{ns-all} where the form of $f$ is
\begin{equation}
    f(R)=f_0 R^n\,,
\end{equation}
with $n$ and $f_0$ are constants. In this case, there are two Noether symmetries  ${\bf X}_0 $ and
\begin{eqnarray}
& & {\bf X}_1 = (2n -3) A \partial_A + B \partial_B + M \partial_M - R \, \partial_R \,. \label{X1-3}
\end{eqnarray}
Then, the corresponding first integrals are
\begin{eqnarray}
& & E_{\mathcal{L}} = 0 \quad \Longleftrightarrow \quad B =  \frac{n M}{ 2n  + (1-n)M R} \left[ \frac{ M'}{ M} \left( \frac{A'}{A} + \frac{M'}{2 M} \right) + (n-1) \frac{ R'}{ R} \left( \frac{A'}{A} + \frac{2 M'}{ M} \right) \right]\,, \label{fint-X0-3}  \\ & & I_1 = \frac{ n f_0 M R^{n-1}}{ \sqrt{A B} } \left[ (2-n) A' + (n-1) (2n-1) A \frac{R'}{R} \right] \,.  \label{fint-X1-3}
\end{eqnarray}
Integration of Eq.\eqref{fint-X1-3} with respect to $A$ gives
\begin{equation}
A(r) = R^{ \frac{ (1-n)(2n-1) }{ 2-n} } \left[ A_0 + \frac{I_1 }{2 f_0 n (2-n)} \int{ R^{ \frac{ (n - 1)(4 n - 5) }{ 2(2 - n)} } \frac{ \sqrt{B}}{M}  dr}  \right]^2\,,  \label{A-3}
\end{equation}
where $A_0$ is an integration constant, and $n \neq 2$.

In order to consider a Schwarzschild-like metric, one has to assume the relation $B(r) = 1/ A(r)$, which causes the following form of $A(r)$,
\begin{equation}
A(r) = R^{ \frac{ (n-1)(2n-1) }{n-2} } \left[ A_0 + \frac{I_1 }{f_0 n (2-n)} \int{  \frac{ R^{ \frac{ 3(n - 1)^2}{ (2 - n)}} }{M}  dr}  \right]\,.  \label{A-4}
\end{equation}
This solution becomes the Schwarzschild solution if $n=1$ and $M(r)= r^2$, in which, one notices that $A_0 = 1$ and $I_1 = 2 f_0 m$ (see~\eqref{r-scalar} and \eqref{fint-X0-3}). Now, by taking $M(r) = r^q$ and $R(r) = R_0 r^{-q}$, we can  find  new exact solutions in power-law $f(R)$ gravity for $A$ that comes from the Eqs~\eqref{r-scalar}, \eqref{fint-X0-3} and \eqref{A-4}. For $n=4$, one finds from these equations that $q=-4/19$, i.e. $M= r^{-4/19},R= R_0 r^{4/19}$ and
\begin{equation}
A(r) = r^{42/19} \left( A_0 + \frac{k}{R_0^3 \, r^{31/19}} \right)\,,  \label{A-5}
\end{equation}
with $A_0 = \frac{361}{992}(8 - 3 R_0), \, k = \frac{19 I_1}{248 f_0 }$ and $R_0 = \frac{72}{23}$.
Furthermore, it follows from Eqs.~\eqref{r-scalar}, \eqref{fint-X0-3} and \eqref{A-4} that we can find other solutions. For example, for $q=2, R_0=1$ and $n=1/2$, we find from \eqref{A-4} that
\begin{equation}
A(r) = \frac{1}{2} - \frac{k}{ r^2}  \qquad {\rm with} \quad  k= \frac{2 I_1}{3 f_0}\,,  \label{A-6}
\end{equation}
is a solution of the field equations. Similarly, for $q= 4/3, R_0=-5$ and $n=5/4$, one finds the following solution
\begin{equation}
A(r) = -k + \frac{9}{2} r^{2/3}  \qquad {\rm with} \quad  k= \frac{8 I_1}{5^{5/4} f_0}\,, \label{A-7}
\end{equation}
whereas for $q= 2/3, R_0=4$ and $n=-1$ one finds
\begin{equation}
A(r) =  \frac{k}{r} - \frac{9}{7} r^{4/3}  \qquad {\rm with} \quad k= \frac{16 I_1}{7 f_0}\,. \label{A-8}
\end{equation}
Finally, if $q= -2/11, R_0=40/13$ and $n=5$, we find the solution
\begin{equation}
A(r) =  r^{24/11} \left( A_0 + \frac{k }{ R_0^4 \, r^{19/11}} \right)  \qquad {\rm with} \quad A_0 = \frac{121}{285} (5 - 2 R_0), \quad  k= \frac{11 I_1}{285 f_0}\,. \label{A-9}
\end{equation}
We should point out that these solutions are obtained by taking $A(r) = 1/B(r)$, which is the well-known Schwarzschild-like form. In our previous study \cite{Bahamonde:2018zcq}, we obtained some new spherically symmetric solutions in power-law $f(R)$ gravity, but these are not in the Schwarzschild-type form, that is, $A(r) \neq 1/B(r)$. To the best of our knowledge, the solutions \eqref{A-4}-\eqref{A-9} are new spherically symmetric solutions in power-law $f(R)$ gravity that have the Schwarzschild form.

\subsubsection{Subcase: $f_{2,GG}(G)\neq0, \ f_{1,RR}(R)\neq 0$ and $f_{1,R}(R)\neq 0$ }
\begin{center}
I) Power-law forms
\end{center}
For this case, one finds that there are two Noether symmetries~ satisfying the Eqs.~\eqref{ns-all}, ${\bf X}_0 $, and
\begin{eqnarray}
& & {\bf X}_1 = (4 p -3) A \partial_A +  B \partial_B + M \partial_M - R \, \partial_R  - 2 G \partial_G \, , \label{X1-4}
\end{eqnarray}
while the form of $f$ becomes
\begin{equation}
    f_1(R)=f_0 \,R^{2p}\,,\quad f_2(G)= f_1 \, G^{p}\,,  \label{f-case-a2}
\end{equation}
with $f_0,f_1$ and $p$ being constants. Here the corresponding Noether integrals are $E_{\mathcal{L}} = 0$ for ${\bf X}_0$, which gives
\begin{eqnarray}
&&\frac{ M' A'}{ M A} + \frac{ M'^2}{2 M^2}  + (2 p - 1)\frac{ R'}{ R} \left( \frac{ A'}{ A} + \frac{ 2 M'}{ M} \right) + \frac{f_1}{2 f_0} (p-1) R^{1 -2 p} G^{p-2} \frac{A' G'}{M A} \left( 4 - \frac{3 M'^2}{M B } \right)  \nonumber \\ && - B \left[ \frac{2}{M} + \left\{ 1- 2 p +  \frac{f_1}{f_0} (1-p)\left(\frac{G}{R^2}\right)^p \right\} \frac{R}{2p} \right]  = 0 \,, \label{fint-X0-4}
\end{eqnarray}
and % \frac{R}{G^2} \left( \frac{G}{R^2} \right)^m
\begin{eqnarray}
 I_1 &=& p \sqrt{\frac{ A }{B}} \Big\{ 2 f_0 M R^{2 p -1} \left[ 2 (1 -p) \frac{A'}{A} + (2 p -1)(4 p -1) \frac{R'}{R} \right]  \nonumber \\& & +  f_1 (p-1) G^{p-1} \left[ \left( 4 - \frac{M'^2}{M B} \right) \left( (4 p - 3) \frac{G'}{G} - \frac{2 A'}{A} \right) - \frac{2 M' A' G'}{B A G} \right] \Big\}  \,,  \label{fint-X1-4}
\end{eqnarray}
for ${\bf X}_1$. The case $p=1/2$ and $f_1=0$ was studied before since it is the General Relativity case.  By setting $I_1=0$ one finds that the $B(r)$ metric coefficient becomes
\begin{eqnarray}
    B(r)&=&\Big[2 M \Big(f_0 G^2 M R^{2 p} \left(2 (p-1) R A'+\left(-8 p^2+6 p-1\right) A R'\right)+4 f_1 (p-1) R^2 A' G^{p+1}\nonumber\\
    && -2 f_1 \left(4 p^2-7 p+3\right) A R^2 G^p G'\Big)\Big]^{-1}\times f_1 (p-1) R^2 G^p M' \left(M' \left((4 p-3) A G'-2 G A'\right)+2 M A' G'\right)\,.
\end{eqnarray}
Now, if one assumes that $R=R_0r^{-l},\, G=G_0 r^{-n},\, M=r^q$ and $A=A_0r^m$ and replace these equations into the remaining field equation, one can directly obtain an exact solution, which reads
\begin{eqnarray}
A(r)&=&A_0\,r^m\,,\quad B(r)=B_0 \, r^{\frac{n-4}{2}}\,,\quad M(r)=r^{n/2}\,,\quad R(r)=R_0\,r^{-n/2}\,,\quad G(r)=G_0 \,r^{-n}\,,
\end{eqnarray}
where $n$ is any parameter and the other constants are 
\begin{eqnarray}
B_0&=&\Big[ n^2 (n-2 m) \left(n^2+2 n m+4 m^2\right)  \left(n^2 \left(64 p^3-80 p^2+20 p+3\right)+8 n m (1-2 p)+4 m^2 (1-4 p)\right)\Big]\times\nonumber\\
&&\Big[16 \Big(n^5 \left(32 p^3-40 p^2+8 p+3\right)+2 n^4 m \left(32 p^3-40 p^2+6 p+3\right)+2 n^3 m^2 \left(64 p^3-80 p^2+12 p+5\right)\nonumber\\
&&-16 n^2 m^3 p-2 \left(K_1-8 m^5\right)\Big)\Big]^{-1}\label{B01}\,,\\
f_1&=&-\frac{f_0 R_0^{2 p-1} G_0^{1-p} \left(n^2+2 n m+4 m^2\right)  \left[ n \left(8 p^2-6 p+1\right)+4 m (p-1)\right]}{2 (p-1) \Big[ n^3 (4 p-3) R_0+2 n^2 m (8 p+3 R_0-10)+16 n m^2 (2 p-1)+16 m^3 \Big]}\,,\label{f11}\\
R_0&=&\Big[4 \Big(2 n^5 p \left(8 p^2-10 p+3\right)-2 n^4 m \left(48 p^3-60 p^2+17 p+1\right)-n^3 m^2 \left(64 p^3-80 p^2+4 p+11\right)\nonumber\\
&&+4 n^2 m^3 (6 p-1) -8 m^5 +K_1 \Big)\Big]\times \Big[ n^2 (2m-n) [ ( 3- 4p )n  + 2 m ] [ ( 16 p^2 - 8p -1) n + 2 ( 4 p -1) m ] \Big]^{-1}\,,\\
G_0&=& \frac{4 m (R_0-2) (2 m -n) \left(n^2 R_0+4 n m+8 m^2\right)}{\left(n^2+2 n m+4 m^2\right)^2}\,,\\
K_1^2&=&\left(n^2+2 n m+4 m^2\right)^2 \Big[ 4 n^6 p^2 \left(8 p^2-10 p+3\right)^2-8 n^5 m p \left(24 p^3-46 p^2+29 p-6\right)\nonumber\\
&&+n^4 m^2 \left(-96 p^3+140 p^2-60 p+7\right)+2 n^3 m^3 \left(32 p^3-24 p^2-2 p+3\right)+n^2 m^4 (8 p-3)-4 n m^5+4 m^6\Big] \,.
\end{eqnarray}
For $n=4$, $B(r)=B_0$ becomes a constant and $M=r^2$. It is also possible to get $A(r)=1/B(r)$ if $n=-2 (m-2)$ and $A_0=B_0^{-1}$ which gives $M=r^{2-m}$ and $B=B_0 r^{-m}$.

\bigskip
\begin{center}
II) Logarithmic form
\end{center}
Additionally to the case described above, there is another similar solution of the Noether's symmetry equations~\eqref{ns-all} which has the Noether's vectors ${\bf X}_0 $ and
\begin{eqnarray}
& & {\bf X}_1 =A \partial_A +  B \partial_B + M \partial_M - R \, \partial_R  -2G\partial_G \, , \label{X1-44}
\end{eqnarray}
and the form of $f$ is
\begin{equation}
    f_1(R)=f_0 \,R^{2}\,,\quad f_2(G)= f_{1}\,G \log (G)\,, 
\end{equation}
where $f_0$ and $f_1$ are constants. The conserved quantity associated to \eqref{X1-44} becomes
\begin{eqnarray}
& & I_1 = \sqrt{\frac{A}{B} } \left[ 6 f_0 M R' + 2 f_1 \frac{A'}{A} \left( \frac{M'^2}{B M} - 4 \right) + f_1 \frac{G'}{G} \left( 4 - \frac{M'^2}{B M} - \frac{2 A' M'}{A B} \right) \right] \, . \label{fint-case-2b}
\end{eqnarray}
For the specific case where $I_1=0$, one gets the following exact solution
\begin{eqnarray}
&&A(r)=A_0r^m\,,\quad B(r)=B_0 r^{q-2}\,,\quad M(r)=r^q\,,\\
&&B_0=-\frac{m^3-m^2 q-4 m q^2+6 q^3\pm K_2}{8 (m+q)}\,,\label{B02}\\
&&f_1=-\frac{f_0 \left(m^5+m^4 q-m^3 q^2+m^2 q^3+26 m q^4+40 q^5\mp K_2 \left(m^2+2 m q-q^2\right)\right)}{8 q^2 \left(-m^3+5 m q^2+4 q^3\right)}\,,\label{f12}\\
&&K_2^2 =  m^6 - 2 m^5 q + 5 m^4 q^2 + 36 m^3 q^3 - 36 m^2 q^4 - 64 m q^5 + 64 q^6\,.
\end{eqnarray}
\subsubsection{Subcase: Pure $f(G)$ gravity ($f_{2,GG}(G)\neq0, \ f_{1,R}(R)=0$)}
\begin{center}
I) Power-law form of $f$
\end{center}
The pure $f(G)$ gravity case admits the following power-law solution for $f$ in the Noether's equations \eqref{ns-all},
\begin{equation}
    f(R,G)= f_0 \, G^p \,,
\end{equation}
in which the Noether symmetries admitted are ${\bf X}_0 $ and
\begin{eqnarray}
& & {\bf X}_1 =  (4 p-3) A \partial_A +  B \partial_B + M \partial_M +\eta^4 (r,R,G,A,B,M) \, \partial_R  - 2G \partial_G \, . \label{X1-4BB-case-3a}
\end{eqnarray}
Here the corresponding first integral for ${\bf X}_1$ is
\begin{eqnarray}
& & I_1 = f_0 p (p-1) G^{p-1}\sqrt{\frac{A}{B}} \left\{ ( 4 p-3) \left( 4 - \frac{M'^2}{B M} \right) \frac{G'}{G} - 2 \frac{A'}{A} \left[ 4  + \frac{M'}{B} \left( \frac{G'}{G} - \frac{M'}{M} \right) \right] \right\}. \label{fint-X1-case-3a}
\end{eqnarray}
When $I_1 = 0$, the first integral \eqref{fint-X1-case-3a} yields
\begin{eqnarray}
& & B = \frac{ M' \left[ (4 p -3) \frac{M' G'}{M G} + \frac{2 A'}{A} \left( \frac{G'}{G} - \frac{M'}{M} \right) \right]}{ 4 \left[ ( 4 p -3) \frac{G'}{G} - \frac{2 A'}{A} \right] } \, .
\end{eqnarray}
Then, we can find the following exact solution:
\begin{equation}
A(r) = A_0 r^m, \, B(r) = B_0 r^{q-2}, \,  M(r)= r^q, \, R(r) = R_0 r^{-q}, \, G(r) = G_0 r^{-2 q}
\end{equation}
where the constant parameters are given by
\begin{eqnarray}
& & q = \frac{ m \left( -1 \pm \sqrt{ 64 p^2 - 48 p + 1} \right)}{ 8 p ( 4 p -3)} \, ,  \\ & & B_0 = \frac{ q^2 ( 4 p q  + 3 m - 3 q) }{ 4 ( 4 p q + m -3 q )}\,,\label{B03} \\ & &
R_0 = \frac{  2 m \left( 5 q^2 + 2 m q - 4 p m q - 4 p q^2 - m^2 \right) }{ q^2 ( 4 p q + 3 m - 3 q )} \, , \\ & & G_0 = - \frac{ 64 p m ( 4 p -3) ( 4 p q + m - 3 q ) }{ ( 4 p q + 3 m - 3 q)^2} \, , 
\end{eqnarray}
with $p \neq 3/4$.

There exists another solution for the Noether's equations that do not constrain the form of $f(G)$, and has the Noether's vectors ${\bf X}_0 $ and
\begin{eqnarray}
& & {\bf X}_1 =  \eta_R(r,R,G,A,B,M) \, \partial_R  \, , \label{X1-4BBB}
\end{eqnarray}
but this solution does not help for finding solutions since the first integrals are identically zero. Furthermore, there is another subcase for the branch $f_{RG}=0$ which is when $f_{2,GG}(G)\neq0$ and $f_{1,RR}(R)=0$. For this subcase, there is only one solution which is the same as \eqref{X1-4BBB} and the form of $f(R,G)=f_0R+f_2(G)$, therefore, this subcase is a generalisation of the above solution. That branch also gives a trivial first integral.

\subsection{Case: $f(R,G) = g_1 (G) + R g_2 (G)$ ($f_{RR} = 0$)  } 
In this case, the condition $f_{RR} = 0$  gives $f(R,G) = g_1 (G) + R g_2 (G)$, and the functions $g_1 (G)$ and $g_2 (G)$ could have different forms that will be study separetely.
\begin{center}
I) Power law form of $f$
\end{center}
The first solution of the Noether's equations~\eqref{ns-all} has the following form of $f$
\begin{equation}
 f(R,G)=R g_1(G)+g_2(G)\,,\quad g_2(G)=f_1 G^\beta \,,\quad   g_1 (G) = f_2 G^{\beta - \frac{1}{2}}\,, 
\end{equation}
 where $\beta$ is a constant, while the Noether's vectors are ${\bf X}_0$ and
 \begin{eqnarray}
    & & {\bf X}_1 = (4 \beta -3) A \partial_A  + B \partial_B + M \partial_M - R \partial_R - 2 G \partial_G \,.
    \end{eqnarray}
Thus, we have the following first integrals for this case
\begin{eqnarray}
   & & \frac{ M' A'}{ M A} + \frac{ M'^2}{2 M^2} + \frac{ A'}{M A G} \left( 4 - \frac{ 3 M'^2}{M B} \right) \left[ ( \beta- \frac{1}{2} ) R' + \left\{ \frac{f_1}{f_2} \beta (\beta -1) \sqrt{G} + ( \beta- \frac{1}{2}) ( \beta- \frac{3}{2}) R  \right\} \frac{G'}{G} \right] \nonumber \\ & &  + ( \beta - \frac{1}{2} ) \frac{G'}{G} \left( \frac{ A' }{ A} + \frac{ 2 M' }{M } \right)  - B \left[ \frac{f_1}{f_2} (1-\beta) \sqrt{G}  - ( \beta - \frac{1}{2} ) R + \frac{2}{M} \right] = 0\,,  \\  I_1 &=& f_2 M G^{ \beta - \frac{3}{2}} \sqrt{ \frac{A}{B}} \Big\{ ( \beta - \frac{1}{2} ) ( 4 \beta - 1) G' + 2 (1-\beta ) G \frac{A'}{A} \nonumber \\ & &  - \frac{2 M' A'}{M A B} \left[ (\beta - 1) R' + \left( \frac{f_1}{f_2} \beta (\beta - 1) \sqrt{G} + ( \beta - \frac{1}{2} ) ( \beta - \frac{3}{2} ) R \right) G^{\beta-1} G' \right]  \nonumber \\ & &  + \frac{1}{M} \left( 4 - \frac{M'^2}{M B} \right) \Big[ ( \beta - \frac{1}{2}) \left( (4 \beta - 3)  R' - R \frac{A'}{A} \right) \nonumber \\ & &    + G^{\beta} \left( \frac{f_1}{f_2} \beta (\beta -1) \sqrt{G} + ( \beta - \frac{1}{2}) ( \beta - \frac{3}{2} ) R \right) \left( (4 \beta -3) \frac{G'}{G} - \frac{2 A'}{A} \right) \Big]  \Big\}\,.
 \end{eqnarray}
One can find the following exact solution for $I_1=0$ and $M=r^2$,
\begin{eqnarray}
A(r)&=&A_0r^{m}\,,\quad B(r)=B_0=\frac{- 16 \left(32 \beta ^3-40 \beta ^2+8 \beta +3\right) - m^3 +2 m^2 + 16 \beta  m \pm K_4}{8 (8 \beta +m-6)}\,,\label{B04}\\
K_{4}^2&=&4096 \beta ^2 \left(8 \beta ^2-10 \beta +3\right)^2+m^6-4 m^5+4 (8 \beta -3) m^4+32 \left(32 \beta ^3-24 \beta ^2-2 \beta +3\right) m^3\nonumber\\
&&-64 \left(96 \beta ^3-140 \beta ^2+60 \beta -7\right) m^2-2048 \beta  \left(24 \beta ^3-46 \beta ^2+29 \beta -6\right) m\,,\\
f_1&=&-\Big[f_2 \Big(64 \left(8 \beta ^3-18 \beta ^2+13 \beta -3\right)+4 B_0 \left(-16 \left(8 \beta ^3-18 \beta ^2+13 \beta -3\right)+(2 \beta -3) m^2+(6-4 \beta ) m\right)\nonumber\\ && +(1-2 \beta ) m^4+8 \left(16 \beta ^3-20 \beta ^2+6 \beta +1\right) m^2+8 \left(32 \beta ^3-104 \beta ^2+94 \beta -27\right) m\Big)\Big]\times \nonumber\\ &&\Big[4(\beta -1) \sqrt{2(1-B_0) (m-2) m}\Big(16 \beta  (3-4 \beta )+m^2-2 m\Big)\Big]^{-1}\,.\label{f14}
\end{eqnarray}

\begin{center}
II) Logarithmic forms
\end{center}
The second solution to \eqref{ns-all} corresponding to this branch has the following logarithmic form for the function,
\begin{equation}
 f(R,G)=Rg_2(G)+g_1(G)\,,\quad g_1(G)=f_0 G \log(G)\,, \quad  g_2 (G) = f_1 \sqrt{G}\,, 
\end{equation}
 while the Noether's vectors are ${\bf X}_0$ and
    \begin{eqnarray}
    & & {\bf X}_1 = A \partial_A  + B \partial_B + M \partial_M - R \partial_R - 2 G \partial_G  \,.
    \end{eqnarray}
For this case, the first integral becomes
\begin{eqnarray}
& & I_1 =  \sqrt{ \frac{A}{B G} } \left\{ \frac{3}{2} f_1 M G' - 2 f_0 \sqrt{G} \frac{A'}{A} \left( 4 - \frac{M'^2}{B M} \right) + \left( 4 - \frac{M'^2}{B M} - \frac{ 2 A' M'}{ A B} \right) \left[ f_1 \frac{R'}{2} + \left( f_0 \sqrt{G} - f_1 \frac{R}{4} \right) \frac{G'}{G} \right]  \right\} \, ,
\end{eqnarray}
and for this case, when $I_1 = 0$, the following exact solution appears 
\begin{eqnarray}
&& A=A_0r^m\,,\quad M=r^2\,,\quad B=B_0=-\frac{m^3-2 m^2-16 m+48\pm K_5}{8 (m+2)}\,,\label{B05}\\
&& f_1=-\frac{4 \sqrt{2} f_0 m \Big[ B_0 (m+6)-m-22 \Big]}{\sqrt{m (m-2)(1- B_0} \left[ 4 B_0+m^2+2 m+28\right]}\,,\label{f15}\\
&& K_5^2 = m^6 - 4 m^5 + 20 m^4 + 288 m^3 - 576 m^2 - 2048 m + 4096 \, .
\end{eqnarray}

\bigskip

\subsection{ Case (iii): $f(R,G) = F_1 (R) + G F_2 (R)$ ($f_{GG} = 0$ )}
The condition of this case is $f_{GG} = 0$ which yields $f(R,G) = F_1 (R) + G F_2 (R)$. For this case we find the following solution of the Noether's equations~\eqref{ns-all},
\begin{eqnarray}
& & {\bf X}_1 = (2 q + 1) A \partial_A  + B \partial_B + M \partial_M - R \partial_R - 2 G \partial_G \,,
\end{eqnarray}
where the function $f$ is given by
\begin{equation}
    f(R,G)=f_{0} R^{2+q}+f_{1} G R^{q}\,,
\end{equation}
where $f_{1},f_{0}$ and $q$ are constants. Then, the corresponding first integrals are
\begin{eqnarray}
& & \left[ (q+2) f_{0} R + q f_{1} \frac{G}{R} \right] \left( \frac{ M' A'}{ M A} + \frac{ M'^2}{2 M^2} \right) + q f_{1} \left( 4 - \frac{ 3 M'^2}{M B} \right) \frac{ A' R'}{M A R}  - \frac{f_{0} B R}{M} \left[ 2 (q+2) - (q+1) M R \right] \nonumber \\ & &  +  \left( \frac{A'}{ A} + \frac{ 2 M' }{M } \right) \left[ \left( f_{0} (q+1) (q+2) +  q f_{1} (q-1) \frac{G}{R^2} \right) R' + q f_{1} \frac{G'}{R} \right] = 0\,,   \\ I_1 &=& q f_{1} M R^q \sqrt{ \frac{A}{B}} \Big\{ \,\, \frac{2 M'}{M R} \left( G - \frac{A' R'}{A B} \right) + (2 q + 3) \frac{G'}{R} - \left[ \frac{f_{0}}{f_{1}} (q+2) R + (q-1) \frac{G}{R} \right] \frac{A'}{A} \nonumber \\ & &  + (2 q +3) \left[ \frac{f_{0}}{q f_{1}} (q+1) (q+2) R + (q-1) \frac{G}{R} \right] \frac{R'}{R} +  \left( 4 - \frac{M'^2}{M B} \right) \left[ (2 q +1) \frac{R'}{R} + \frac{A'}{A} \right] \Big\}\, .
\end{eqnarray}
When $I_1=0$, one finds the following solution
\begin{eqnarray}
& & A(r)= A_0r^{m}\,, \,\, M(r) = r^2,  \,\, B(r)=B_0=\frac{-m^3+2 m^2+8 m (q+2)-16 \left(4 q^3+14 q^2+12 q+3\right) \pm K_6 }{8 (m+4 q+2)}\,,\label{B06}\\
& & f_{1} = \frac{f_{0} \left( m^2 +2 m  -4 B_0 +4 \right)^2 \Big[ 4 B_0+m^2 (q+1)+2 m \left(2 q^2+5 q+1\right)+4 \left(4 q^2+12 q+7\right)\Big] }{8 m q \Big[-16 B_0^2+B_0 \left[m^3+4 m^2 (q+1)+4 m (2 q+5)-32 (q-1)\right] - (m+4) \left[m^2+4 m (q+2)-8 q+4\right] \Big]} \,  .  \quad \label{f16} \\ 
& & K_6^2 = 16 (m-2) (m+4 q+2) \left[ m^2 (2 q+3)+8 m (q+1)-4 \left(8 q^3+28 q^2+26 q+7\right)\right] \nonumber\\
&& \qquad \quad + \Big[ m^3-2 m^2-8 m (q+2)+16 \left(4 q^3+14 q^2+12 q+3\right)\Big]^2\,.
\end{eqnarray}

\subsection{$f_{RG}\neq0,\, f_{RR}\neq0$ and $f_{GG}\neq0$}
The last branch is when $f_{RG}\neq0,\, f_{RR}\neq0$ and $f_{GG}\neq0$ which has two subcases that will be studied separately.
\subsubsection{$f_{GG} f_{RR}\neq f_{RG}^2$}
If $f_{GG} f_{RR}\neq f_{RG}^2$, one finds that there are two possible forms of $f$ satisying the Noether's equations~\eqref{ns-all}. The first one is given by 
\begin{equation}
 f(R,G)=   R^{\beta} h\left(\frac{G}{R^2}\right)\,,
\end{equation}
were $h$ is any arbitrary function of the quantity $G/R^2$. This model has the Noether's vector $\bf{X}_0$ and
\begin{eqnarray}
{\bf X}_1 &=&A(3-2\beta)\partial_A-B\partial_B -M\partial_M+R\partial_R+2G\partial_G\,.\label{X1is2}
\end{eqnarray}
There is another possible solution for $f(R,G)$ in this case which has the following form
\begin{equation}
 f(R,G)=  R^2 h\left(\frac{G}{R^2}\right)+f_1 G \log (R)\,,\label{eqa}
\end{equation}
where again $h$ is any arbitrary function of $G/R^2$. This solution has the same form of the Noether's vector  given in \eqref{X1is2} with $\beta=2$. 

For the case~\eqref{eqa}, it is possible to obtain a general form of solution given by
\begin{eqnarray}\label{ABC}
&&A(r)=A_0r^m\,,\quad  B(r)=B_0 =\frac{\pm K_7-m^3+2 m^2+16 m-48}{8 (m+2)}\,,\quad M=r^2\,,\label{B07}\\
&& \frac{G}{R^2}=\frac{4 (m-2) m (m+2) \left(\mp K_7+m^3-2 m^2-8 m+64\right)}{\left(\mp K_7+3 m^3+6 m^2+64\right)^2}\,,\\
&&K_7^2=m^6-4 m^5+20 m^4+288 m^3-576 m^2-2048 m+4096\,,
\end{eqnarray}
and the remaining field equation depends only on the form of the function $h(G/R^2)$, namely
\begin{eqnarray}
    &&8 f_1 m (m+2) \Big(2 K_6^2\mp K_6 \left(m^4+10 m^3+20 m^2+8 m+256\right)+m^7+6 m^6+40 m^4+784 m^3 \nonumber\\
    &&+1856 m^2+1024 m+8192\Big) -4 (m-2) m (m+2) \Big(-K_6^2\mp 4 K_6 (3 m+22) m+m^6+8 m^5+52 m^4-112 m^3\nonumber\\
    &&-128 m^2+4608 m+4096\Big) h'\left(\frac{G}{R^2}\right)+\left(\pm K_6+m^3+10 m^2+80 m+64\right) \left(\mp K_6+3 m^3+6 m^2+64\right)^2 h\left(\frac{G}{R^2}\right)=0\,.\nonumber\\\label{h0}
\end{eqnarray}
Thus, if one assumes a specific form of $h$, one can easily solve the above algebraic equation. For example, if $h(G/R^2)=h_0 \log (G/R^2)$, one easily gets that
\begin{eqnarray}
h_0&=&-\Big[8 f_1 m (m+2) \left(K_6+m^3-2 m^2-8 m+64\right) \Big(2 K_6^2-K_6 \left(m^4+10 m^3+20 m^2+8 m+256\right)+m^7+6 m^6+40 m^4\nonumber\\
&&+784 m^3+1856 m^2+1024 m+8192\Big)\Big]\times\Big[\left(K_6+m^3+10 m^2+80 m+64\right) \Big(-\left(\pm K_6+m^3-2 m^2-8 m+64\right)\times \nonumber\\
&&\left(\pm K_6+3 m^3+6 m^2+64\right)^2+\left(K_6+m^3-2 m^2-8 m+64\right) \left(-K_6+3 m^3+6 m^2+64\right)^2 \log \left(G/R^2\right)\Big)\Big]^{-1}\,, \label{h0-2}
\end{eqnarray}
where $G/R^2$ becomes a constant. One can find many more solutions like \eqref{ABC} by assuming a form of $h=h(G/R^2)$, replacing it in \eqref{h0} and solving for the constants. 
\subsubsection{$f_{GG} f_{RR}= f_{RG}^2$}
There is an additional branch having different Noether's symmetries, which is the one where $f_{GG} f_{RR}= f_{RG}^2$. For simplicity, for this branch, we will assume that the function is separable as
\begin{equation}
    f(R,G)=f_1(R)f_2(G)\,,
\end{equation}
which gives us the following differential equations
\begin{equation}
    \frac{f_2(G) f''_2(G)}{f'_2(G)^2}-\frac{f'_1(R)^2}{f_1(R) f''_1(R)}=0\,.
\end{equation}
Since the first term depends only on $G$ and the second only on $R$, one finds that
\begin{equation}
    -\frac{f'_1(R)^2}{f_1(R) f''_1(R)}=p\,,\quad  \frac{f_2(G) f''_2(G)}{f'_2(G)^2}=-p\,.
\end{equation}
The above equations can be easily solved, yielding ($p\neq-1$)
\begin{eqnarray}
f_1(R)&=&f_0 (-f_1 p+p R+R)^{\frac{p}{p+1}}\,,\quad 
f_2(G)=f_2 (-f_3+G p+G)^{\frac{1}{p+1}}\,,\quad p\neq -1\,,\label{powerlaw}
\end{eqnarray}
and for $p=-1$,
\begin{eqnarray}
f_1(R)&=&f_0 e^{f_3 R}\,,\quad f_2(G)=e^{f_1 G}\,,\quad p=-1\,,\label{exponential}
\end{eqnarray}
where $f_i$ $(i=1,..3)$ are constants. By replacing these forms into the Noether's equations~\eqref{ns-all}, and after doing some computations one finds that the form \eqref{powerlaw} has three Noether's vector, $\bf{X}_0$, 
\begin{eqnarray}
 {\bf X}_1 &=&   \eta_R(r,R,G,A,B,M) \partial_R +\Big(\frac{f_3-G (p+1)}{f_1 p-(p+1) R}\Big)\eta_R(r,R,G,A,B,M) \partial_G  \,,\label{X1is}\\
{\bf X}_2 &=&A\partial_A-\frac{B (p+1)}{p-1}\partial_B -\frac{M (p+1)}{p-1}\partial_M+\eta_R(r,R,G,A,B,M)\partial_R\nonumber\\
&+&\Big[\frac{ \eta_R(r,R,G,A,B,M)}{R}+\frac{ (p+1)}{p-1}\Big]G\partial_G\,, \quad f_1=f_3=0\,.
\end{eqnarray}
The exponential form of $f$ given in \eqref{exponential} (case $p=-1$), has two Noether's vectors, $\bf{X}_0$ and ${\bf X}_1$ given by \eqref{X1is} with $p=-1$. The Noether's vector $\bf{X}_1$ gives a zero Noether's charge for both $p=-1$ and $p\neq-1$ cases.
The first integral for ${\bf X}_1$ becomes $I_1 = 0$, and for ${\bf X}_2$ it has the form
\begin{eqnarray}
 I_2& =& f_0 M \sqrt{ \frac{A}{B} }\left( \frac{G}{R} \right)^{\frac{1}{p+1}} \Big\{ \frac{p (p+2)}{(p-1)} \frac{A'}{A} + \frac{2 p (p+1)}{(p-1)} \frac{M'}{M} + \frac{p(p+3)}{p^2 -1} \left( \frac{R'}{R} - \frac{G'}{G} \right) - \frac{p R A'}{(p-1) M A} \left( 4 - \frac{M'^2}{B M}  \right) \nonumber \\ & &  + \frac{R}{M G} \left[ \frac{p}{p+1} \left( 4 - \frac{M'^2}{B M} \right) + \frac{ 2 A' M'}{(p-1) A B} \right] \left( \frac{R'}{R} - \frac{G'}{G} \right) \Big\} \, ,
\end{eqnarray}
where $f_1 = f_3 = 0$ which yields $f(R,G) = f_0 (p+1) R^{\frac{p}{p+1}} G^{ \frac{1}{p+1}}$.

\section{Conclusions}
\label{sec:V}
In this work, we classified $f(R,G)$ gravity, a generalization of $f(R)$ containing a Gauss-Bonnet scalar in the action, that are invariant under point transformations and thus possess Noether symmetries. We found more than ten different forms of $f(R,G)$ which present some Noether symmetries that vary from power-law to logarithmic, involving different couplings. 
Some of these models  in Friedmann-Lemaitre-Robertson-Walker universe were already known to reproduce the late-time acceleration of the universe, without knowing that they have such symmetries~\cite{Odintsov2019}. This means that the conserved quantities of these models could be immediately related to observables.

Apart from that, we used these symmetries to solve the field equations of $f(R,G)$ for these models, and found exact spherically symmetric solutions. The obtained first integrals related with the Noether symmetries include mostly two independent equations for five unknown quantities $A,B,M,R$ and $G$. It has to be noted here that $R$ and $G$ are depended on $A, B$ and $M$ due to the definitions of $R$ by \eqref{r-scalar} and $G$ by \eqref{gaussbonnet}. Therefore, once we set one of the metric coefficients $A, B$ and $M$, and then the number of unknowns reduce to two. In order to reduce the number of these quantities, one can also use some suitable assumptions. For studying  the spherically symmetry, there are two common settings which are the gauges $M(r)= r^2$ and $B(r)= 1/A(r)$. Choosing one of these gauges we have obtained some exact spherically symmetric solutions for several forms of the function $f(R,G)$ which are represented in Table \ref{Table1}. 

In the context of galactic dynamics, it is represented that some of the modified gravity models such as $f(R)$ gravity can describe the flat rotation curves of the galaxies 
\cite{Capozziello:2012ie,Boehmer:2007kx}. The static and spherically symmetric metric given by \eqref{metric} could be relevant to obtain some important quantities for the galactic dynamics. For instance, the $g_{tt}$ component of the metric tensor, which is the function $A(r)$ in this study, determines the tangential velocity of a test particle by using the constants of motion that is defined via geodesic motions. Furthermore, it is interesting to note that the tangential velocity is independent of the form of metric function $g_{rr} = B(r)$. For the constant tangential velocity regions of galaxies, the metric tensor component $g_{tt} = A(r)$ can be written as $A(r) = A_0 r^m$, where the power $m$ is a constant related with the tangential velocity ( see  ref. \cite{Boehmer:2007kx} for details). In this study, we obtained that type of metric functions for some $f(R,G)$ gravity models summarized in Table \ref{Table1}. This issue will be considered in future works.

\begin{table}[h]
\centering
\resizebox{15cm}{!}{\begin{tabular}{|c |c | c |c|} 
\hline
 $\mathbf{f(R,G)}$ & $\mathbf{A(r)}$ & $\mathbf{B(r)}$ & $\mathbf{M(r)}$  \\ [0.5ex] \hline 
 $f_0 R^4$ & $r^{42/19} \left[ -\frac{361}{713}  + \frac{k}{( 72/23)^3 \, r^{31/19}} \right]$   & \multirow{5}*{$1/A(r)$}  & $r^{-4/19}$   \\ [0.9ex] \cline{1-2}\cline{4-4}
 $f_0 R^{1/2}$& $\displaystyle\frac{1}{2} -\displaystyle \frac{k}{ r^2}$ &  & $r^2$ \\ [0.9ex] \cline{1-2}\cline{4-4}
 $f_0 R^{5/4}$ & $-k + \frac{9}{2} r^{2/3} $  & & $r^{4/3}$ \\ [0.9ex] \cline{1-2}\cline{4-4}
 $f_0 R^{-1}$ & $ \frac{k}{r} - \frac{9}{7} r^{4/3}$ & & $r^{2/3}$ \\ [0.9ex] \cline{1-2}\cline{4-4}
 $f_0 R^{5}$& $r^{24/11} \left[ -\frac{121}{247} + \frac{k }{ (40/13)^4 \, r^{19/11}} \right]$ & &  $r^{-2/11}$\\ [0.9ex]\hline
 $f_0R^{2p}+f_1^{(1)} G^{p}$& $A_0r^{m}$ & $B_0^{(1)}r^{(n-4)/2}$ & $r^{n/2}$ \\ [0.9ex]\hline
$f_0R^2+f_1^{(2)} G \log(G)$ & $A_0r^{m}$ & $B_0^{(2)}r^{q-2}$& $r^{q}$\\ [0.9ex]\hline
$f_0 G^p$ & $A_0r^{m}$ & $B_0^{(3)}r^{\frac{ m \left( -1 \pm \sqrt{ 64 p^2 - 48 p + 1} \right)}{ 8 p ( 4 p -3)}-2}$& $r^{\frac{ m \left( -1 \pm \sqrt{ 64 p^2 - 48 p + 1} \right)}{ 8 p ( 4 p -3)}}$\\ [0.9ex]\hline 
$f_1 G^{\beta}+f_2 R \, G^{\beta-\frac{1}{2}}$ & $A_0r^{m}$ & $B_0^{(4)}$& $r^2$\\ [0.9ex]\hline
$f_0 G \log G+f_1^{(5)} R \sqrt{G}$& $A_0 r^m$& $B_0^{(5)}$& $r^2$\\ [0.9ex]\hline
$f_0 R^{2+q}+f_1^{(6)} G\, R^{q}$& $A_0r^{m}$ & $B_0^{(6)}$ & $r^2$\\ [0.9ex]\hline
$R^2 h\left(\frac{G}{R^2}\right)+f_1 G \log (R)$ &$A_0r^m$& $B_0^{(7)}$& $r^2$\\ [0.9ex]\hline
  \end{tabular}}
\caption{Exact solutions in $f(R,G)$ where the metric is $ds^2=-A(r)dt^2+B(r)dr^2+M(r)d\Omega^2$. For simplicity, we have added the labels $(i)$ with $i=1,..,7$ in some constants appearing in the solutions. For $B_0^{(1)}, f_1^{(1)}, B_0^{(2)}, f_1^{(2)}, B_0^{(3)}, B_0^{(4)}, B_0^{(5)},f_1^{(5)},B_0^{(6)},f_1^{(6)}$ and $B_0^{(7)}$, see Eqs.~\eqref{B01},~\eqref{f11},~\eqref{B02},~\eqref{f12},~\eqref{B03},~\eqref{B04},~\eqref{B05},~\eqref{f15},~\eqref{B06},~\eqref{f16} and \eqref{B07}, respectively. The form of $ h\left(\frac{G}{R^2}\right)$ is any function satisfying Eq.~\eqref{h0}. }\label{Table1}
 \end{table}\newpage
\section*{Acknowledgements}
S.B. is supported by Mobilitas Pluss No. MOBJD423 by the Estonian government. This article is based upon work from CANTATA COST (European Cooperation in Science andTechnology) action CA15117, EU Framework Programme Horizon 2020.
\newpage

\appendix
\section{Noether's symmetry equations }\label{appendix}
The Noether symmetry condition \eqref{ngs-eq} applied to the Lagrangian \eqref{lagr} for the metric \eqref{metric} gives $60$ partial differential equations, that read
\begin{eqnarray}
& & f_{R} \xi_{,_A} = 0, \quad f_{R} \xi_{,_B} = 0, \quad f_{RR} \xi_{,_A} = 0, \quad f_{GR} \xi_{,_A} = 0, \quad f_{GG} \xi_{,_A} = 0, \quad  f_{RR} \xi_{,_B} = 0, \quad f_{GR} \xi_{,_B} = 0, \quad f_{GG} \xi_{,_B} = 0, \nonumber \\& & f_{R} \xi_{,_M} = 0, \quad f_{GR} \xi_{,_M} = 0, \quad f_{GG} \xi_{,_M} = 0, \quad f_{RR} \xi_{,_R} = 0, \quad f_{GR} \xi_{,_R} = 0, \quad f_{GG} \xi_{,_R} = 0, \nonumber \\ & & f_{RR} \xi_{,_G} = 0, \quad f_{GR} \xi_{,_G} = 0, \quad f_{GG} \xi_{,_G} = 0, \quad f_{GR} \eta^1_{,_B} = 0, \quad f_{GG} \eta^1_{,_B} = 0, \quad f_{GR} \eta^3_{,_B} = 0, \quad f_{GG} \eta^3_{,_B} = 0, \nonumber \\ & &
f_{GR}\eta^{1}_{,M}=0\,, \quad f_{GG}\eta^{1}_{,M}=0\,, \quad f_{GR}\eta^{3}_{,A}=0\,,\quad f_{GG}\eta^{3}_{,A}=0\,, \quad  f_{GR}\eta^{1}_{,G}+f_{GG}\eta^{1}_{,R}=0\,.\,,\nonumber\\ &&
f_{GR} \eta^1_{,_R} = 0, \quad f_{GG} \eta^1_{,_G} = 0, \quad f_{GR} \eta^3_{,_R} = 0, \quad f_{GG} \eta^3_{,_G} = 0, \quad  f_{GG} \eta^3_{,_R} + f_{GR} \eta^3_{,_G} = 0, \quad  f_{GR} \eta^4_{,_A} + f_{GG} \eta^5_{,_A} = 0,  \nonumber \\ & &  f_{GR} \eta^4_{,_B} + f_{GG} \eta^5_{,_B} = 0, \quad f_{GR} \eta^4_{,_M} + f_{GG} \eta^5_{,_M} = 0, \quad M f_{RR} \eta^1_{,_B} + 2 A f_{RR} \eta^3_{,_B} = 0,  \quad  A B f_{R} \xi_{,_G} + 2 f_{GG} \eta^1_{,_r} = 0,  \nonumber  \\ & &  A B \left( f_{R} \xi_{,_R} + 4 M f_{RR} \xi_{,_M} \right) + 2 f_{GR} \eta^1_{,_r} = 0, \quad M B f_{R} \xi_{,_G} + 2 f_{GG} \eta^3_{,_r} = 0, \quad  M B \left( f_{R} \xi_{,_R} + M f_{RR} \xi_{,_M} \right) + 2 f_{GR} \eta^3_{,_r} = 0, \nonumber \\ & &  \left( M f_{RR} + 4 f_{GR} \right) \eta^1_{,_R} + 2 A f_{RR} \eta^3_{,_R} = 0, \quad \left( M f_{GR} + 4 f_{GG} \right) \eta^1_{,_G} + 2 A f_{GR} \eta^3_{,_G} = 0, \,\, f_{GR} \eta^4_{,_r} + f_{GG} \eta^5_{,_r} = 0, \nonumber \\& &  \left( M f_{RR} + 4 f_{GR} \right) \eta^1_{,_G} + \left( M f_{GR} + 4 f_{GG} \right) \eta^1_{,_R} + 2 A \left( f_{GR} \eta^3_{,_R} + f_{RR} \eta^3_{,_G} \right) = 0, \quad V \xi_{,_B} + K_{,_B} = 0, \nonumber \\& & f_{R} \eta^3_{,_r} + \left( M f_{RR} + 4 f_{GR} \right) \eta^4_{,_r} + \left( M f_{GR} + 4 f_{GG} \right) \eta^5_{,_r} - \sqrt{ A B} \left( V \xi_{,_A} + K_{,_A} \right) = 0, \nonumber \\ & & f_{R} \left( \frac{1}{A} \eta^1_{,_r} + \frac{1}{M} \eta^3_{,_r} \right) + 2 \left( f_{RR} \eta^4_{,_r} + f_{GR} \eta^5_{,_r} \right)  - \sqrt{ \frac{B}{A} } \left( V \xi_{,_M} + K_{,_M} \right) = 0,  \nonumber \\& & 2 A f_{RR} \eta^3_{,_r} + \left( M f_{RR} + 4 f_{GR} \right) \eta^1_{,_r}  - \sqrt{ A B} \left( V \xi_{,_R} + K_{,_R} \right) = 0, \nonumber \\ & &  2 A f_{GR} \eta^3_{,_r} + \left( M f_{RG} + 4 f_{GG} \right) \eta^1_{,_r}  - \sqrt{ A B} \left( V \xi_{,_G} + K_{,_G} \right) = 0, \nonumber \\ & & f_{R} \eta^3_{,_A} + \left( M f_{RR} + 4 f_{GR} \right) \eta^4_{,_A} + \left( M f_{GR} + 4 f_{GG} \right) \eta^5_{,_A} = 0, \nonumber \\ & & f_{R} \eta^3_{,_B} + \left( M f_{RR} + 4 f_{GR} \right) \eta^4_{,_B} + \left( M f_{GR} + 4 f_{GG} \right) \eta^5_{,_B} = 0,  \label{ns-all} \\ & &  2 A \left( f_{RR} \eta^4_{,_B} + f_{GR} \eta^5_{,_B} \right) + f_{R} \eta^1_{,_B} + \frac{A}{M} f_{R} \eta^3_{,_B} = 0, \nonumber \\ & &  -\frac{f_{R}}{2} \left( \frac{\eta^1}{A} + \frac{\eta^2}{B} \right) + f_{RR} \eta^4 + f_{GR} \eta^5 + f_{R} \left( \eta^1_{,_A} + \frac{A}{M} \eta^3_{,_A} + \eta^3_{,_M} - \xi_{,_r} \right) + \left( M f_{RR} + 4 f_{GR} \right) \eta^4_{,_M} \nonumber \\& & \qquad \qquad +  \left( M f_{GR} + 4 f_{GG} \right) \eta^5_{,_M} + 2 A \left(  f_{RR} \eta^4_{,_A} + f_{GR} \eta^5_{,_A} \right) = 0, \nonumber \\ & & \left( M f_{RR} + 4 f_{GR} \right) \left[ -\frac{1}{2} \left( \frac{\eta^1}{A} + \frac{\eta^2}{B} \right) + \eta^1_{,_A} + \eta^4_{,_R} - \xi_{,_r} \right] + f_{RR} \eta^3 + \left( M f_{RRR} + 4 f_{GRR} \right) \eta^4 + \left( M f_{RRG} + 4 f_{GRG} \right) \eta^5 \nonumber \\ & & \qquad \qquad + f_{R} \eta^3_{,_R} + 2 A f_{RR} \eta^3_{,_A} + \left( M f_{GR} + 4 f_{GG} \right) \eta^5_{,_R} = 0,  \nonumber \\ & & \left( M f_{GR} + 4 f_{GG} \right) \left[ -\frac{1}{2} \left( \frac{\eta^1}{A} + \frac{\eta^2}{B} \right) + \eta^1_{,_A} + \eta^5_{,_G} - \xi_{,_r} \right] + f_{GR} \eta^3 + \left( M f_{RGR} + 4 f_{GGR} \right) \eta^4 + \left( M f_{RGG} + 4 f_{GGG} \right) \eta^5 \nonumber \\ & & \qquad \qquad + f_{R} \eta^3_{,_G} + 2 A f_{GR} \eta^3_{,_A} + \left( M f_{RR} + 4 f_{GR} \right) \eta^4_{,_G} = 0, \nonumber \\ & & f_{R} \left[   \frac{1}{2 M} \left( \eta^1 - \frac{A}{B} \eta^2 \right) - \frac{A}{M^2} \eta^3 + 2 \eta^1_{,_M} + \frac{A}{M} \left( 2 \eta^3_{,_M} - \xi_{,_r} \right) \right] + \frac{A}{M} \left( f_{RR} \eta^4 + f_{GR} \eta^5 \right) + 4 A \left( f_{RR} \eta^4_{,_M} + f_{GR} \eta^5_{,_M} \right) = 0, \nonumber \\ & & f_{RR} \left(  \frac{\eta^1}{2 A} - \frac{\eta^2}{2 B} + \eta^3_{,_M} + \eta^4_{,_R} - \xi_{,_r} \right) + f_{RRR} \eta^4 + f_{RRG} \eta^5 + \frac{f_{R}}{2} \left( \frac{1}{A} \eta^1{,_R} + \frac{1}{M} \eta^3_{,_R} \right) \nonumber \\ & & \qquad \qquad + f_{GR} \eta^5_{,_R} + \frac{1}{2 A} \left( M f_{RR} + 4 f_{GR} \right) \eta^1_{,_M}  = 0,  \nonumber \\ & & f_{GR} \left(  \frac{\eta^1}{2 A} - \frac{\eta^2}{2 B} + \eta^3_{,_M} + \eta^5_{,_G} - \xi_{,_r} \right) + f_{RGR} \eta^4 + f_{RGG} \eta^5 + \frac{f_{R}}{2} \left( \frac{1}{A} \eta^1{,_G} + \frac{1}{M} \eta^3_{,_G} \right) \nonumber \\ & & \qquad \qquad + f_{RR} \eta^4_{,_G} + \frac{1}{2 A} \left( M f_{GR} + 4 f_{GG} \right) \eta^1_{,_M}  = 0, \nonumber \\ & & f_{GR} \left[ -\left( \frac{\eta^1}{2 A} +\frac{3 \eta^2}{2 B} + \frac{\eta^3}{M} \right) + \eta^1_{,_A} + 2 \eta^3_{,_M} + \eta^4_{,_R} - 3\xi_{,_r}  \right] + f_{GRR} \eta^4 + f_{GRG} \eta^5 + f_{GG} \eta^5_{,_R} = 0,  \nonumber \\ & & f_{GG} \left[ -\left( \frac{\eta^1}{2 A} +\frac{3 \eta^2}{2 B} + \frac{\eta^3}{M} \right) + \eta^1_{,_A} + 2 \eta^3_{,_M} + \eta^5_{,_G} - 3 \xi_{,_r}  \right] + f_{GGR} \eta^4 + f_{GGG} \eta^5 + f_{GR} \eta^4_{,_G} = 0, \nonumber \\ & & V_{,_A} \eta^1 + V_{,_B} \eta^2 + V_{,_M} \eta^3 + V_{,_R} \eta^4 + V_{,_G} \eta^5 + V \xi_{,_r} + K_{,_r} = 0, \nonumber
\end{eqnarray}
where $V \equiv \sqrt{A B} \left[ M (G f_G - f) + (M R -2) f_R \right]$.

%\bibliographystyle{Style}
%\bibliography{bibtele}

 \bibliographystyle{utphys}
 \bibliography{references}

\end{document}